\documentclass[a4paper,english,british,a4paper,fleqn,usenatbib]{mnras}
\usepackage[T1]{fontenc}
\usepackage[latin9]{inputenc}
\usepackage{units}
\usepackage{url}
\usepackage{amssymb}
\usepackage{graphicx}
\usepackage{esint}
\usepackage[authoryear]{natbib}

\makeatletter

\usepackage[T1]{fontenc}
\usepackage{ae,aecompl}

\title[The imprint of combustion on a superburst]{The imprint of carbon combustion on a superburst from the accreting neutron star 4U\,1636--536}
\author[L. Keek, A. Cumming, Z. Wolf, et al.]{L. Keek,$^1$\thanks{E-mail: l.keek@gatech.edu} A. Cumming,$^2$ Z. Wolf,$^1$ D.R. Ballantyne,$^1$ V.F. Suleimanov,$^{3,4}$\newauthor E. Kuulkers,$^5$ T.E. Strohmayer$^6$ \\
$^1$Center for Relativistic Astrophysics, School of Physics, Georgia Institute of Technology, 837 State Street, Atlanta, GA 30332-0430, USA\\
$^2$Department of Physics, McGill University, 3600 rue University, Montreal, QC, H3A 2T8, Canada\\
$^3$Institut f\"ur Astronomie und Astrophysik, Kepler Center for Astro and Particle Physics, Universit\"at T\"ubingen, Sand 1, 72076 T\"ubingen,\\ \ Germany\\
$^4$Kazan (Volga region) Federal University, Kremlevskaja str., 18, Kazan 420008, Russia\\
$^5$European Space Astronomy Centre (ESA/ESAC), Science Operations Department, 28691 Villanueva de la Cañada, Madrid, Spain\\
$^6$X-ray Astrophysics Lab, Astrophysics Science Division, NASA's Goddard Space Flight Center, Greenbelt, MD 20771, USA
}

\date{Accepted XXX. Received YYY; in original form ZZZ}
\pubyear{2015}

\makeatother

\usepackage{babel}
\begin{document}
\label{firstpage}
\pagerange{\pageref{firstpage}--\pageref{lastpage}}
\maketitle
\begin{abstract}
Superbursts are hours-long X-ray flares attributed to the thermonuclear
runaway burning of carbon-rich material in the envelope of accreting
neutron stars. By studying the details of the X-ray light curve, properties
of carbon combustion can be determined. In particular, we show that
the shape of the rise of the light curve is set by the the slope of
the temperature profile left behind by the carbon flame. We analyse
\emph{RXTE}/PCA observations of 4U~1636--536 and separate the direct
neutron star emission from evolving photoionized reflection and persistent
spectral components. This procedure results in the highest quality
light curve ever produced for the superburst rise and peak, and interesting
behaviour is found in the tail. The rising light curve between 100
and 1000 seconds is inconsistent with the idea that the fuel burned
locally and instantaneously everywhere, as assumed in some previous
models. By fitting improved cooling models, we measure for the first
time the radial temperature profile of the superbursting layer. We
find $d\ln T/d\ln P\approx\nicefrac{1}{4}$. Furthermore, $20\,\%$
of the fuel may be left unburned. This gives a new constraint on models
of carbon burning and propagation in superbursts.\end{abstract}

\section{Introduction}

In the past decade rare long (hours--day) X-ray flares have been observed
from neutron stars that accrete from a companion star in low-mass
X-ray binaries \citep{Cornelisse2000}. In contrast to the shorter
($10$--$100\,\mathrm{s}$) Type I X-ray bursts powered by hydrogen/helium
flashes \citep[e.g.,][]{Lewin1993}, these so-called superbursts are
thought to arise from runaway carbon fusion in the ashes of hydrogen
and helium burning (\citealt{Strohmayer2002,Cumming2001}). 

\citet{2004CummingMacBeth} assumed that after superburst ignition
the carbon would rapidly undergo complete burning to iron group elements,
and modelled the superburst light curve by numerically following the
subsequent cooling of the neutron star envelope. These cooling models
were fit to superburst light curves to derive the depth and energy-content
of the burning layer \citep{Cumming2006}. Although they are a good
match to the tail of superbursts, these models failed to reproduce
the rise of the observed light curves. In particular, the cooling
model light curves predict that the luminosity decreases between 100
to 1000 seconds after onset of the superburst, whereas the observed
light curves are increasing in luminosity at that time. With the same
assumption of complete burning, \citet{Weinberg2007} also found declining
light curves at 100\textendash 1000 seconds. In contrast, \citet{Keek2011}
carried out fully self-consistent simulations of superbursts that
followed the carbon burning and energy deposition in detail. They
instead found a rising luminosity during this phase of the superburst
in some models (where the precursor did not dominate the rise). 

This earlier work indicates that the phase of the superburst light
curve prior to the peak at $\sim1000$ seconds is sensitive to the
details of how the carbon combustion deposits energy in the neutron
star envelope. The physics of how the carbon burning propagates is
complex and may not be accurately captured by current 1D models. Rather
than explicitly treating the hydrodynamic process of flame spreading,
these models resolve the burning layers in the radial direction and
include a 1D approximation to turbulence. In this paper we derive
a useful constraint on such modelling by using the observed light
curve to constrain the radial temperature profile left behind by carbon
burning. This may inform other applications of carbon flames, such
as the study of detonation and deflagration in Type Ia supernovae
\citep[e.g.,][]{Woosley2011}.

Most of the current $24$ (candidate) superbursts (e.g., \citealt{Keek2012};
see \citealt{Negoro2012} and \citealt{Serino2014ATel} for recent
detections) are detected with instruments that have limited sensitivity
\citep[e.g.,][]{Zand2003,Keek2008}. Furthermore, the rise is often
not observed due to gaps in the data from Earth occultation or passage
through the South-Atlantic Anomaly \citep[e.g.,][]{Zand2004}. The
two highest quality observations were performed with the Proportional
Counter Array (PCA; \citealt{Jahoda2006}) on-board the \emph{Rossi
X-ray Timing Explorer} (\textit{RXTE}; \citealt{Bradt1993}). The
first was a bright superburst from 4U~1820--30 \citep{Strohmayer2002,Ballantyne2004}.
Its luminosity quickly reached the Eddington limit, which impedes
the study of the intrinsic rise \citep[e.g.,][]{Zand2010}. The second
PCA superburst observation was from 4U~1636--536 \citep{Strohmayer2002a,2004Kuulkers}.
It was not Eddington-limited, and at present it provides the only
detailed observation of a superburst rise. Moreover, it is the only
(super)burst for which spectral analyses were able to separate the
direct burst emission from an evolving persistent component as well
as photoionized reflection off the accretion disc \citep{Keek2014sb1,Keek2014sb2}.
Reflection refers to scattering of burst emission off the disc, which
introduces features in the spectrum, such as a strong Fe~K$\alpha$
emission line, that depend on the ionization state of the disc \citep{Ballantyne2004models}.

In this paper we fit improved cooling models to the superburst light
curve of 4U~1636--536. First we investigate the robustness of the
light curve derived from the observations, in particular with respect
to the influence of deviations of the neutron star's spectrum from
a blackbody (Section~\ref{sec:Reconstructing-the-Observed}). Subsequently,
we create improved cooling models to constrain properties of the superbursting
layer, including the slope of its temperature profile (Section~\ref{sec:Cooling-Models}).
Comparison of the superburst tail to the best-fitting model light
curve reveals interesting, but not entirely understood, behaviour
(Section~\ref{sec:tail}). Finally, we discuss how carbon combustion
produces the inferred temperature gradient (Section~\ref{sec:Discussion}).

\section{Reconstructing the observed light curve}

\label{sec:Reconstructing-the-Observed}

For comparison to superburst cooling models, we set out to determine
the most reliable light curve of neutron-star emission by separating
out emission from and reflection off the accretion disc. In previous
analyses, a blackbody was used to model the thermal spectrum of the
neutron star \citep{2004Kuulkers,Keek2014sb1,Keek2014sb2}. A neutron
star atmosphere spectrum is, however, expected to exhibit deviations
from a blackbody due to, e.g., electron scattering which influence
the blackbody parameters obtained in a spectral analysis \citep{Suleimanov2010}.
To investigate how this affects the inferred net superburst light
curve, we repeat the analysis procedure of \citet{Keek2014sb2} with
neutron star atmosphere spectra instead of a blackbody.

\subsection{Observations and spectral models}

\label{sub:Observations-and-Spectral}

We reuse the PCA source spectra, instrumental background spectra,
and response matrices that were created for \citet{Keek2014sb2} \citep[see also][]{Keek2014sb1}.
The spectra are collected in $64\,\mathrm{s}$ time intervals, and
we consider the $3-20\,\mathrm{keV}$ energy range in our analysis.
The superburst spans $4$ \emph{RXTE} orbits. As the signal is weaker
in the final orbits, we use a single spectrum covering orbit $3$.
Orbit $4$ is omitted, because the reflection features cannot be detected,
and, therefore, the reflection signal cannot be separated from direct
burst emission (\citealt{Keek2014sb2}). Spectral fitting is performed
with XSPEC version 12.8.1 \citep{Arnaud1996}, and $1\,\sigma$ errors
in the fit parameters are derived using a change in goodness of fit
of $\Delta\chi^{2}=1$.

The spectra are described using $3$ components: persistent emission
from the accretion disc, direct thermal emission from the neutron
star, and reflection of the thermal emission off the disc. The persistent
part is modelled by a power law with an exponential high-energy cut-off
at energy $E_{\mathrm{cutoff}}$. For the other two components \citet{Keek2014sb2}
employed a blackbody with temperature $kT$ and a model of a reflected
blackbody \citep{Ballantyne2004models} at the same temperature, as
a function of the disc's ionization parameter $\xi$ (in units of
\foreignlanguage{english}{$\mathrm{erg\,s^{-1}\,cm}$} in this paper).
We refer to this model as \textbf{BB}. The spectral model further
includes the photoelectric absorption model \texttt{vphabs} with an
equivalent hydrogen column $N_{\mathrm{H}}$ \citep[c.f.,][]{Pandel2008}
and smoothing of the reflection component due to Doppler broadening
and relativistic effects using the \texttt{rdblur} model \citep{Fabian1989}.

As an alternative for the blackbody, we use neutron star atmosphere
spectra based on the models by \citet{Suleimanov2012} with a surface
gravity of $\log g=14.3$ ($g$ in units of $\mathrm{cm\,s^{-2}}$).
Instead of the blackbody temperature, these models are a function
of the ratio of the luminosity to the Eddington limited luminosity:
$L/L_{\mathrm{Edd}}$. We create two grids of atmosphere spectra each
containing $100$ values of the Eddington ratio, equidistant on a
logarithmic scale, in the range $10^{-3}\leq L/L_{\mathrm{Edd}}\leq1.08$.
These grids are optimized for the analysis of the superburst, as they
have improved resolution at lower values of $L/L_{\mathrm{Edd}}$
compared to \citet{Suleimanov2010}. One grid is for solar composition
(model \textbf{S1}), and the other has a $100$ times lower metal
content (model \textbf{S001}). During the spectral fits with each
grid, we fit for $L/L_{\mathrm{Edd}}$ and the normalization. For
both grids we calculate photoionized reflection spectra using the
code of \citet{Ballantyne2002code} using the same procedure as \citet{Ballantyne2004models}.

\subsection{Spectral analysis}

\label{sub:Spectral-Analysis}

\begin{figure*}
\includegraphics{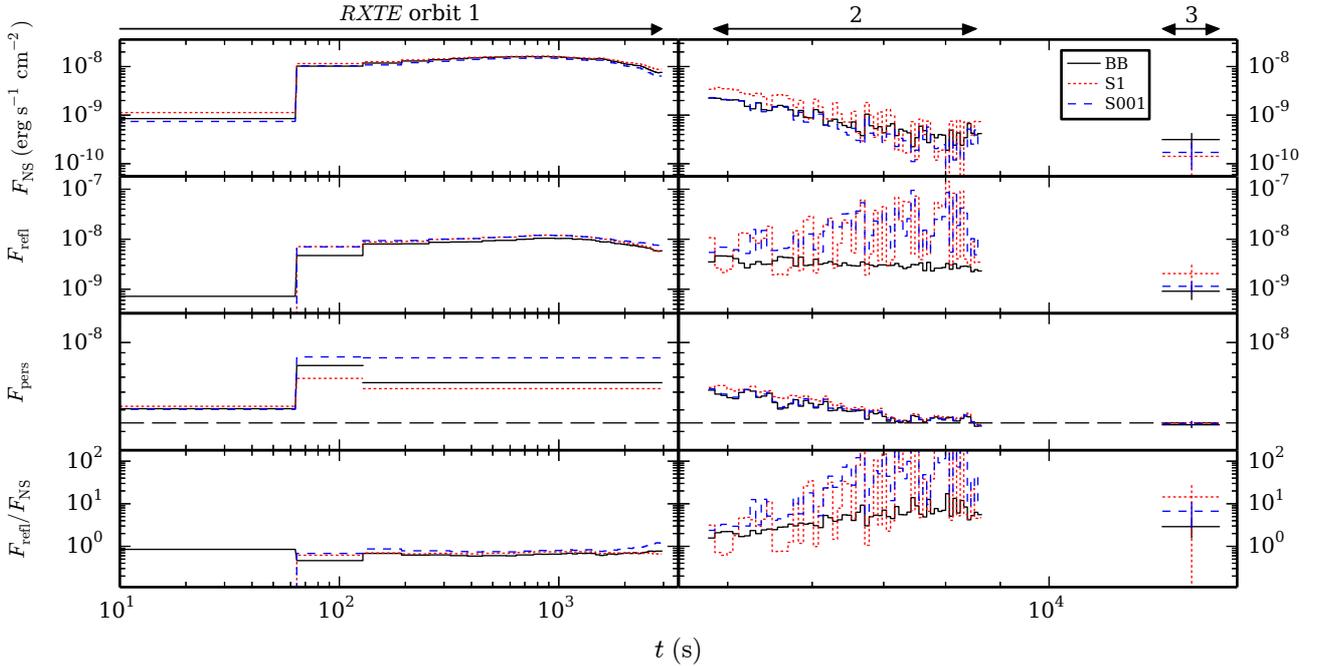}

\caption{\label{fig:compare-flux}For $3$ spectral models, flux as a function
of time, $t$, for the direct neutron star emission ($F_{\mathrm{NS}}$),
reflection ($F_{\mathrm{refl}}$), and persistent flux ($F_{\mathrm{pers}}$;
the long-dashed line is the pre-superburst value). At the bottom we
show the reflection fraction $F_{\mathrm{refl}}/F_{\mathrm{NS}}$.
$F_{\mathrm{NS}}$ is modelled with a blackbody in \textbf{BB} and
by a neutron star atmosphere in \textbf{S1} and \textbf{S001}. S1
assumes solar metallicity, whereas S001 uses a 100 times lower metalicity.
The left and right panels each have a different horizontal scale.
The $1\sigma$ uncertainty is only indicated for the last data point.
Excellent fits are obtained around the peak (left panel), where $F_{\mathrm{NS}}$
for the 3 models agrees at all times within $17\%$. Issues in the
tail appear as large \emph{unphysical} variations of $F_{\mathrm{refl}}$
for S1 and S001 (right panel; see also Section~\ref{sub:Interpretation-of-the}).}
\end{figure*}
Because of the limited data quality, \citet{Keek2014sb2} found that
certain parameters need to be constrained during the spectral fits.
In particular, in the first \emph{RXTE} orbit (excluding the first
$128\,\mathrm{s}$, which contain the pre-superburst flux and the
precursor burst; see Figure~\ref{fig:compare-flux}) all spectra
are fit simultaneously, and the normalizations of the power law and
the direct neutron star component as well as $N_{\mathrm{H}}$ are
tied between the spectra. The power-law index and $E_{\mathrm{cutoff}}$
are fixed to the pre-superburst values \citep{Keek2014sb1}. In orbits
$2$ and $3$, $E_{\mathrm{cutoff}}$ is free, but the normalization
of the neutron star component is fixed to the value obtained for the
first orbit. This procedure yielded a good fit with model BB \citep{Keek2014sb2},
and repeating it with model S1 gives a similar goodness of fit. It
does not, however, result in a good fit for S001 in the first orbit.
To obtain a good fit with S001, we allow $E_{\mathrm{cutoff}}$ to
vary in the first orbit, keeping its value tied between the spectra
in the first orbit. The best-fitting value, $E_{\mathrm{cutoff}}=6.17\pm0.07$,
is significantly larger than the pre-superburst value of $E_{\mathrm{cutoff}}=4.80\pm0.06$
\citep{Keek2014sb1}. The $\chi_{\nu}^{2}$ values obtained with S1
and S001 have a similar distribution as the values for model BB \citep{Keek2014sb2}. 

S1 and S001 yield parameter values that are over-all quantitatively
consistent with model BB. In particular, the weighted mean of $\log\xi$
in the first orbit is within $2.8\%$ of the BB value of $\log\xi=3.384\pm0.009$,
and in the second orbit all models exhibit a transition to a lower
value of $\log\xi$. Furthermore, the Eddington ratio of the atmosphere
models corresponds to an effective temperature of the neutron star
photosphere that is lower than the measured blackbody temperature
\citep{Suleimanov2010}. The ratio of the two is referred to as the
colour correction factor, $f_{\mathrm{c}}$. For model S1, we find
$f_{\mathrm{c}}=1.464\pm0.007$ when the Eddington ratio peaks at
$L/L_{\mathrm{Edd}}=0.499\pm0.007$, and it drops to a mean value
of $f_{\mathrm{c}}=1.295\pm0.010$ in the first $512\,\mathrm{s}$
of the second orbit where $L/L_{\mathrm{Edd}}=0.101\pm0.002$. Similarly,
for model S001 the Eddington ratio peaks at $L/L_{\mathrm{Edd}}=0.441\pm0.006$
with $f_{\mathrm{c}}=1.511\pm0.008$, and maintains a comparable value
of $f_{\mathrm{c}}=1.483\pm0.011$ while the Eddington ratio decreases
to $L/L_{\mathrm{Edd}}=0.0604\pm0.0013$ at the start of the second
orbit. In both cases the behaviour is consistent with the theoretical
predictions for the considered values of $L/L_{\mathrm{Edd}}$: $f_{\mathrm{c}}$
is close to constant for a metal-poor atmosphere, but it exhibits
larger variations for solar composition \citep{Suleimanov2010,Suleimanov2012}.
The deviations of the spectrum from a blackbody are, therefore, too
small for the data to discriminate between S1 and S001, and the models
simply behave as predicted.

The neutron star fluxes, $F_{\mathrm{NS}}$, are consistent for BB,
S1, and S001: in the first orbit within $17\%$ (Figure~\ref{fig:compare-flux}).
Because S001 has a higher $E_{\mathrm{cutoff}}$, its persistent flux,
$F_{\mathrm{pers}}$, is larger by $29\%$ in the first orbit. In
the following orbits S001's $F_{\mathrm{pers}}$ is consistent with
BB and S1. Furthermore, in the second orbit the reflected neutron
star flux, $F_{\mathrm{refl}}$, of both S1 and S001 exhibits strong
variations. For some spectra the fit prefers low values of $L/L_{\mathrm{Edd}}$,
which require large bolometric corrections and produce unrealistically
large values for $F_{\mathrm{refl}}$ and for the reflection fraction
$F_{\mathrm{refl}}/F_{\mathrm{NS}}$. For other spectra, the fits
find more reasonable values of $F_{\mathrm{refl}}$, and this `track'
is closer to the fluxes and reflection fractions obtained for BB (Figure~\ref{fig:compare-flux}).
We attribute these issues to the increasingly poor data quality in
the superburst tail. 

Because of the similarity of $F_{\mathrm{NS}}$ for all models around
the peak, and the issues of S1 and S001 in the tail, we choose model
BB for comparison with theoretical light curves. We calculate the
net superburst luminosity from $F_{\mathrm{NS}}$ assuming isotropic
emission by the neutron star at a distance of $6.0\pm0.5\,\mathrm{kpc}$
\citep{Galloway2008catalog}. We do not include the distance error
in the uncertainty of the luminosity data points, but it is included
as a systematic uncertainty when we fit cooling models (Section~\ref{sec:Cooling-Models}).
Because of the mentioned issues in the tail of the superburst, we
first focus on the rise and the peak of the superburst in the first
orbit. In Section~\ref{sec:tail} we look at the tail in more detail.

\section{Cooling models for the superburst light curve}

\label{sec:Cooling-Models}

In this section, we make models of the superburst light curve by assuming
that the nuclear burning occurs rapidly after ignition, leaving behind
a temperature profile that we then take as an initial condition for
the subsequent cooling. We show that the light curve slope during
the rising phase at $100$--$100\,{\rm s}$ constrains the temperature
profile in the layer set by carbon burning.

\subsection{Details of the models}

\label{sub:Details-of-the}

To calculate the cooling of the heated layer, we follow the approach
of \citet{2004CummingMacBeth} and \citet{Cumming2006}. The main
differences from those calculations are in the choice of initial temperature
profile for the cooling, a correction to the radiative opacities used
in those papers, and a more accurate outer boundary condition.

To follow the cooling of the layer, we integrate\footnote{The code used in this paper is available at \url{https://github.com/andrewcumming/burstcool}.}
the heat equation 
\begin{equation}
c_{P}\frac{\partial T}{\partial t}=\frac{\partial F}{\partial y}-\epsilon_{\nu};\hspace{1cm}F=\frac{4acT^{3}}{3\kappa}\frac{\partial T}{\partial y},
\end{equation}
where $F$ is the heat flux, $\epsilon_{\nu}$ the neutrino emissivity,
$T$ the temperature, $y$ the column depth, $\kappa$ the opacity,
$a$ the radiation constant, and $c$ the speed of light. We use the
method of lines, in which we difference the spatial part of these
equations on a grid in column depth with a constant spacing $\Delta x=\Delta\ln y$,
and then integrate the stiff set of ordinary differential equations
that result forward in time using an implicit integrator. The inner
zone is set to be a factor of $1000$ deeper in column than the ignition
depth. We typically set the outer zone in our time-dependent calculation
at a column depth $y_{t}=10^{8}\ {\rm g\ cm^{-2}}$, which is a small
fraction of the typical fuel column ($\lesssim10^{-3}$). 
We adopt as default neutron star parameters $M=1.4\ M_{\odot}$ and
$R=12\ {\rm km}$ giving a gravity $g_{14}=g/10^{14}\ {\rm cm\ s^{-2}}=1.60$
and redshift $1+z=(1-2GM/c^{2}R)^{-1/2}=1.24$, where $g=(GM/R^{2})(1+z)$.
This choice of $g$ is slightly smaller than what was used for the
atmosphere spectra (Section~\ref{sub:Observations-and-Spectral}),
but there is no problem with consistency, because we will fit the
cooling models to the light curve from spectral model BB. The calculation
is Newtonian in that redshift variations across the thin layer are
ignored. Instead, we set the gravity to be a constant and then redshift
the time and flux to infinity to compare to observations.

The microphysics input is as follows. For the equation of state of
the electrons, we use the fits of \citet{Paczynski1983}, except for
the electron heat capacity which is a modified version of the \citet{Paczynski1983}
formula that has the correct limits for non-degenerate and degenerate
electrons \citep{Schatz2003ApJ}. The electron Fermi energy is calculated
as needed according to \citet{Chabrier1998}. Coulomb corrections
for the ions are calculated using the results of \citet{Potekhin2000}
(see their Equation.~17). The free-free and electron scattering radiative
opacities are calculated following \citet{1999Schatz}. When forming
the total opacity, we include an additional factor from \citet{Potekhin2001}
to account for the fact that Rosseland mean opacities are not additive
\citep[see discussion in][]{Stevens2014}. 
The thermal conductivity is calculated by implementing the fitting
formulas of \citet{Potekhin1999}. Neutrino emissivities due to the
plasma, pair, and bremsstrahlung processes are included using the
fitting formulae from \citet{Schinder1987} and \citet{Haensel1996}.

The boundary condition $F(T)$ at $y=y_{t}$ is determined from a
grid of constant flux atmospheres which are integrated separately.
The key region is the ``sensitivity strip'' at column depths $\sim10^{7}$--$10^{8}\ {\rm g\ cm^{-2}}$,
which is assumed to consist of $^{56}$Ni in our default model (matching
the composition of the cooling layer). 
The layers $y<10^{7}\ {\rm g\ cm^{-2}}$ are pure helium (the composition
at low density does not affect the $F$--$T$ relation, being out
of the sensitivity strip). 

\begin{figure}
\includegraphics[width=1\columnwidth]{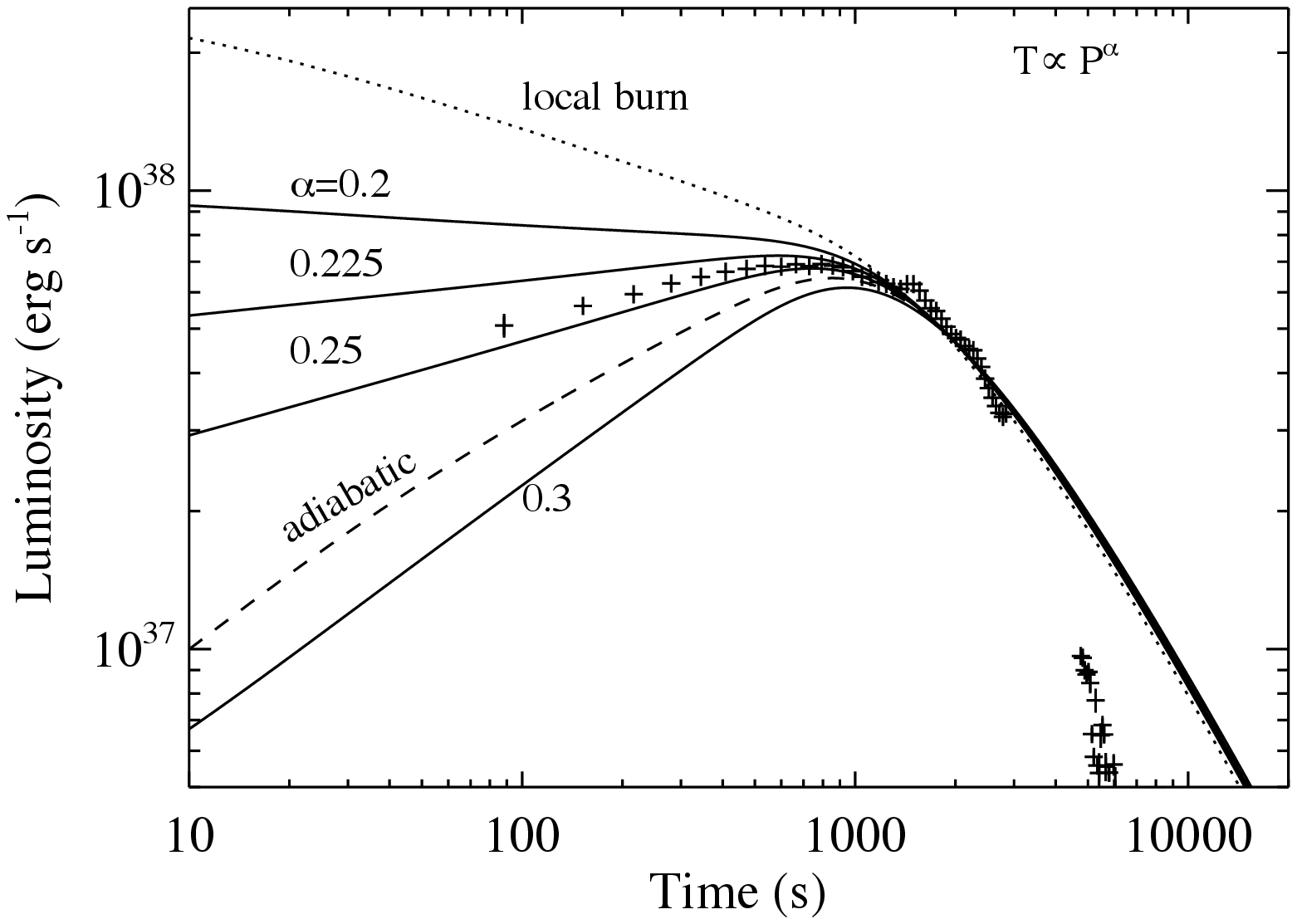} \includegraphics[width=1\columnwidth]{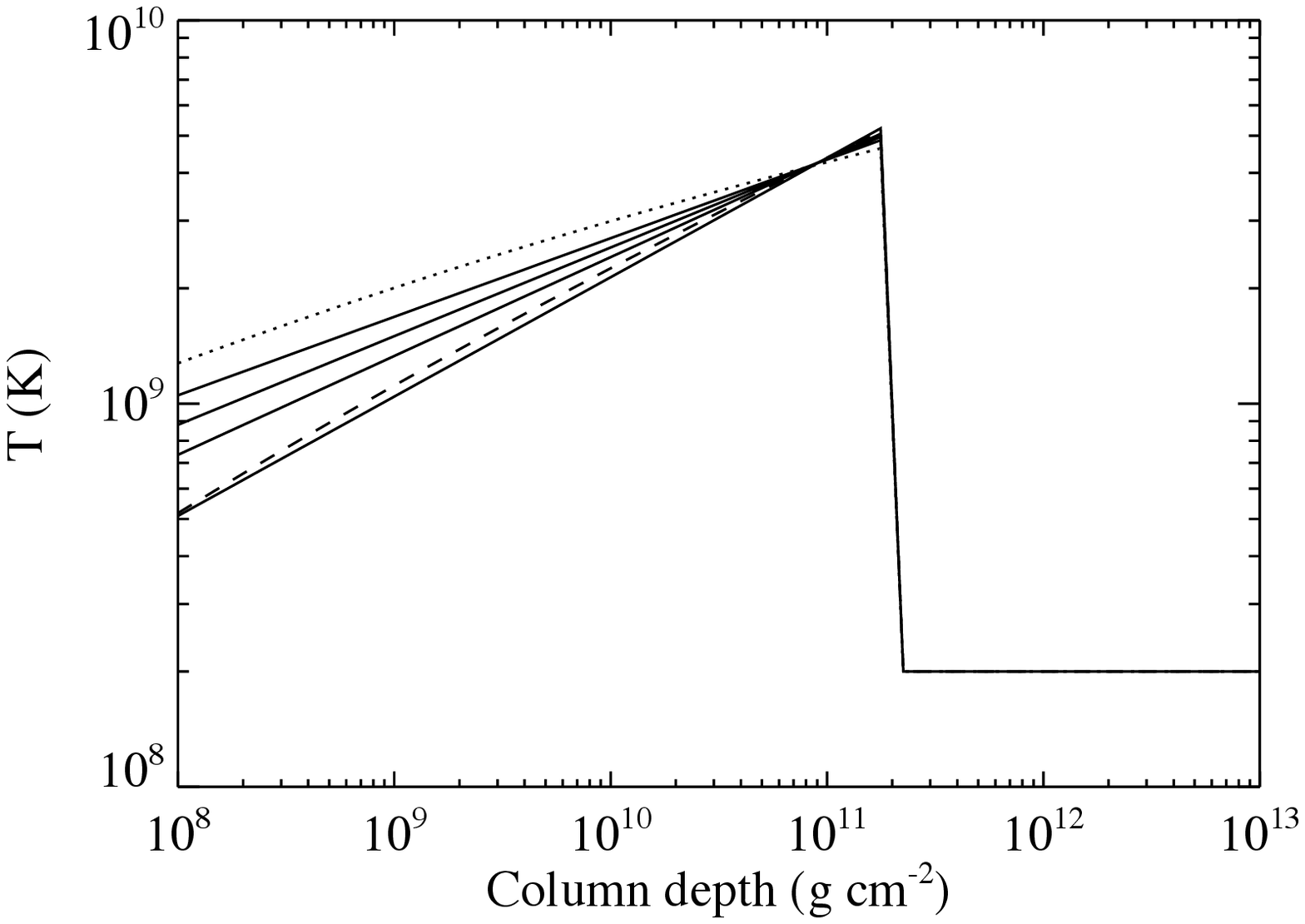}\caption{\label{fig:alpha} Model light curves (top panel) with the same total
energy release and ignition depth, but different initial temperature
profiles (bottom panel). The solid curves have $T\propto P^{\alpha}$
with (top to bottom) $\alpha=0.2$, $0.225$, $0.25$ and $0.3$.
The dot-dashed curve assumes local deposition of energy ($\alpha\approx1/8$
as in \citealt{2004CummingMacBeth}); the dashed curve has an adiabatic
profile $d\ln T/d\ln P=\nabla_{{\rm ad}}$. For all models the ignition
column is $2\times10^{11}\,{\rm g\,cm^{-2}}$ and energy release $E_{18}=0.25$.
Crosses are the data points of the observation of 4U\,1636--536,
where time starts at the onset of the precursor ($t=72\,\mathrm{s}$
in Figure~\ref{fig:compare-flux}).}
\end{figure}

\subsection{The initial temperature profile}

The starting temperature profile for the cooling is set to be a power
law with depth, $T_{f}(y)=T_{b}(y/y_{b})^{\alpha}$. The normalization
$T_{b}$ is chosen such that the average energy input is a specified
value $E_{\mathrm{nuc}}=E_{18}10^{18}\ {\rm erg\ g^{-1}}$, 
\begin{equation}
\int_{y_{t}}^{y_{b}}dy\int_{T_{i}}^{T_{f}(y)}c_{P}dT=yE_{{\rm nuc}}.
\end{equation}
Since $T_{f}\gg T_{i}$, the light curves are not very sensitive to
the choice of $T_{i}$; for simplicity we set $T_{i}=2\times10^{8}\ {\rm K}$,
constant with depth. 

The effect of the choice of the slope of the temperature profile $\alpha$
on the light curve is shown in Figure~\ref{fig:alpha}. The light
curves shown have the same total energy release and depth, but different
choices of $\alpha$. We also show two other choices of initial temperature
profile. One is to assume no heat transport, and burn the fuel locally
and instantaneously. This was the assumption of \citet{2004CummingMacBeth}
and gives a profile with $\alpha\approx1/8$. The second is an adiabatic
initial profile, with $d\ln T/d\ln P=\nabla_{\mathrm{ad}}$ at each
depth. This could result if convection is able to efficiently mix
the burning layer.

\begin{figure}
\includegraphics[width=1\columnwidth]{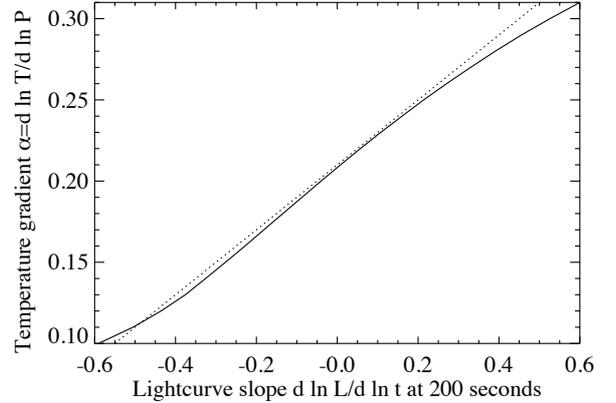} \caption{The slope of the light curve at 200 seconds ($d\ln L/d\ln t$) as
a function of the temperature slope in the layer at the onset of cooling
$T\propto P^{\alpha}$. The dotted line is the analytically-motivated
fit $d\ln L/d\ln t=5(\alpha-0.21)$. \label{fig:slope_alpha}}
\end{figure}

Figure~\ref{fig:alpha} shows that different choices of $\alpha$
result in different light curve slopes at times $100$--$1000\ {\rm s}$
before the light curve peak. Smaller values of $\alpha\lesssim0.21$
lead to a luminosity that falls with time during this phase; larger
values of $\alpha$ give a luminosity that increases with time. We
see that a measurement of the slope of the cooling curve at early
times is a direct measure of the slope of the temperature profile
in the layer immediately following burning. This is explicitly shown
in Figure~\ref{fig:slope_alpha} which shows the temperature slope
$\alpha$ as a function of the slope of the light curve measured at
$200$ seconds. 

The form of the relation between $\alpha$ and $d\ln L/d\ln t$ shown
in Figure~\ref{fig:slope_alpha} can be understood using the analytic
estimates for the early time light curve from Appendix A of \citet{Cumming2006}.
Assuming constant opacity for simplicity, they modelled the early
time light curve by writing the flux $F\propto T^{4}/y$, where $y$
is the depth that the cooling wave has reached at a time $t$. This
depth is given by setting the thermal time-scale to the current time
$t=t_{{\rm therm}}\propto y^{7/4}/T^{2}$. With a temperature profile
of the form $T\propto y^{\alpha}$, we find $d\ln L/d\ln t=-4(4\alpha-1)/(8\alpha-7)$.
This qualitatively agrees with Figure~\ref{fig:slope_alpha}, but
is not accurate in detail as we might expect from a simple constant
opacity estimate. We find that $d\ln L/d\ln t=5(\alpha-0.21)$ is
a good approximation to the numerical results, shown in Figure~\ref{fig:slope_alpha}
as a dotted line.

\subsection{Comparison to the data}

\label{sub:Comparison-to-the}

The data for the 4U\,1636--536 light curve is plotted in Figure~\ref{fig:alpha}.
We take the start time $t=0$ at the onset of the precursor burst
\citep[e.g.,][]{Keek2014sb1}, which corresponds to $t=72\,\mathrm{s}$
in Figure~\ref{fig:compare-flux}. As described by \citet{Cumming2006},
the total energy release is constrained by the luminosity of the early
part of the light curve (before the peak at $\approx10^{3}$ seconds),
whereas the depth of the layer is constrained by the cooling time-scale
associated with the tail of the burst (after $10^{3}$ seconds). We
find that an energy release $E_{18}\approx0.25$ and a depth $y\approx2\times10^{11}\ {\rm g\ cm^{-2}}$
give a good fit to the data. In addition, as can be seen in Figure~\ref{fig:alpha},
the data strongly constrain the temperature profile to have $\alpha=0.25$.
The observed slope $d\ln L/d\ln t$ is positive, ruling out the local
burning temperature profile, but also not steep enough to match the
adiabatic profile.

The curves shown in Figure~\ref{fig:alpha} are for a particular
choice of neutron star mass and radius, composition of the burning
products, and distance (Section~\ref{sub:Details-of-the}). We investigate
the robustness of our conclusions to these parameters. We use the
\texttt{emcee} Markov Chain Monte Carlo (MCMC) code \citep{ForemanMackey2013emcee}
to explore the parameter space. In doing so, we fit the data for $t<1400\,{\rm s}$
only. This includes the peak of the light curve, but excludes most
of the decay. As can be seen in Figure~\ref{fig:alpha}, the observed
luminosity shows abrupt changes of slope during the decay phase which
can bias the fits, since there are many data points in the decay part
of the light curve. By including the data in the rise and peak only
we are able to obtain good fits to the data ($\chi_{\nu}^{2}\approx1$)
and we still obtain a tight constraint on the allowed column depth.

We find that the derived values of $E_{18}$, $y$ and especially
$\alpha$ are not very sensitive to the uncertainties in the other
parameters. For a fixed distance, we find that changing the heavy
element composition, for example from Fe to Si, or the mass and radius
of the neutron star within typical values, e.g. $1.2$ to $2.0M_{\odot}$
and $10$\textemdash $13\,{\rm km}$ changes the derived energy and
column depth by $\approx10\,\%$, with much smaller changes in $\alpha$.
Changing the distance has a larger effect on the derived energy and
column depth \citep{Cumming2006}. For example, with a prior on distance
from $4$ to $8\,{\rm kpc}$, we find a range of $E_{18}\approx0.24$\textemdash $0.32$
and $y\approx1.5$\textemdash $4\times10^{11}\,{\rm g\,cm^{-2}}$.
The distribution of allowed slopes is still rather narrow, with $\alpha=0.242\pm0.004$.

\section{The tail of the superburst}

\label{sec:tail}

\begin{figure*}
\includegraphics{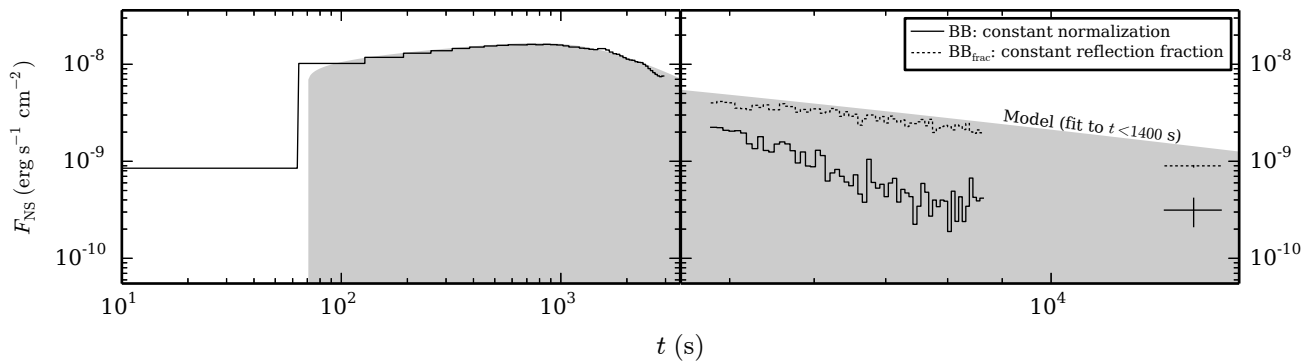}

\caption{\label{fig:compare-model}Comparison of two derived superburst light
curves to the cooling model that fits best to the first $1400\,\mathrm{s}$
(top of the shaded region; see Section~\ref{sec:Cooling-Models}).
\textbf{BB} assumes a constant normalization of the spectral component
that describes the thermal emission of the neutron star, whereas spectral
model \textbf{BB$_{\mathbf{frac}}$} assumes the reflection fraction
to be constant. The left and right panels and their scale are the
same as in Figure~\ref{fig:compare-flux}.}
\end{figure*}
Although the cooling models are in excellent agreement with the rise
and peak of the observed light curve, in the tail $F_{\mathrm{NS}}$
decreases much faster than predicted (see the corresponding luminosity
in Figure~\ref{fig:alpha}). As the normalization of the neutron
star spectral component is kept fixed in the tail, the behaviour of
the flux is purely due to the temperature evolution. This implies
exceptionally fast cooling of the neutron star atmosphere. 

As an alternative assumption for the normalization of the BB blackbody
component in the tail, we fix the reflection fraction to the mean
value in the first orbit: $F_{\mathrm{refl}}/F_{\mathrm{NS}}=0.665\pm0.012$
(model $\mathbf{BB_{frac}}$). The physical implication of a constant
reflection fraction is that the accretion geometry is unchanged during
the superburst. The goodness of fit for $\mathrm{BB_{\mathrm{frac}}}$
is similar as the other models. This produces a slower decay that
closely follows the trend of the cooling model (Figure~\ref{fig:compare-model}).
To quantify how well $\mathrm{BB_{\mathrm{frac}}}$ matches the cooling
model, we take the ratio of $F_{\mathrm{NS}}$ to the model flux.
In orbit 2, the weighted mean is $0.858\pm0.009$. This implies that
the actual reflection fraction in the tail is $0.43\pm0.02$. A decrease
of the reflection fraction may be expected if the disc was puffed
up by X-ray heating in the peak, and returned to a flatter geometry
as the illumination was reduced in the tail \citep[e.g.,][]{Ballantyne2005,lapidus85mnras}.

Model $\mathrm{BB_{\mathrm{frac}}}$ has a similarly fast decrease
of the blackbody temperature as BB. The slower decline of the flux,
therefore, implies that the temperature decrease is partially compensated
by an increase in the normalization. In the final $1024\,\mathrm{s}$
of the second orbit, the weighted mean of the ratio of $F_{\mathrm{NS}}$
for BB$_{\mathrm{frac}}$ and BB is $5.2\pm0.2$. As the blackbody
normalization is proportional to the apparent emitting area, this
corresponds to a radius increase with respect to the peak by a factor
$2.27\pm0.04$. The appearance of radius expansion in this part of
the superburst is puzzling, and in Section~\ref{sub:Interpretation-of-the}
we further discuss the challenges of the interpretation of the superburst
tail.

\section{Discussion}

\label{sec:Discussion}

We first discuss the main result from our analysis: constraints on
the superburst parameters from cooling models, including the temperature
profile created by the carbon flame. Next we compare the inferred
ignition column to the observational constraints on the accreted column.
Finally, we discuss the robustness of the observed light curve and
in particular the issues with the interpretation of the spectra in
the superburst tail.

\subsection{Retracing the carbon flame with cooling models}

Carbon combustion has been studied extensively in the context of Type
Ia supernovae, in particular to investigate whether carbon flames
propagate as a detonation and/or a deflagration \citep[e.g.,][]{Woosley2011}.
Superbursts provide a unique opportunity to study carbon flames, as
they occur close to the surface of the star, without the thermonuclear
event disrupting it. \citet{Weinberg2007} calculated the propagation
of a detonation through the carbon layer. At depths $y\lesssim y_{{\rm det}}\approx2\times10^{11}\ {\rm g\ cm^{-2}}$
the time-scale to burn carbon is longer than a dynamical time-scale,
so that the detonation ceases and the shock runs ahead of the burning.
The column depth that we infer for the 2001 4U~1636--536 superburst
is about this depth, suggesting that a detonation did not likely occur
in this superburst.

Our fits with cooling models find a temperature profile with power-law
slope $\alpha\simeq\nicefrac{1}{4}$, whereas complete local burning
produces $\alpha=\nicefrac{1}{8}$. One way in which a temperature
profile with $\alpha>\nicefrac{1}{8}$ could come about is if the
burning becomes more and more incomplete at low densities. \citet{Weinberg2007}
pointed out that the shallow superburst light curves imply a steeper
temperature gradient than would result from a complete burn. They
suggested that perhaps the convective deflagration leaves behind an
increasing amount of unburned fuel as it moves to lower densities.
Since local instantaneous burning produces a temperature $T_{f}\propto E_{18}^{1/2}y^{1/8}$
\citep{2004CummingMacBeth}, $T_{f}\propto P^{\alpha}$ implies $E_{18}\propto P^{(8\alpha-1)/4}$.
For $\alpha=\nicefrac{1}{4}$ this is $E_{18}\propto P^{1/4}$. If
the fuel layer initially has a homogeneous composition, this describes
the fraction of the fuel that is burned as a function of depth. For
example, assuming that at the ignition depth $100\%$ was burned,
we estimate that only $50\%$ was burned at an order of magnitude
shallower depth. Integrating over all depths, as much as $20\%$ by
mass of the fuel may be left unburned.

Sound waves or a shock generated by carbon ignition have been suggested
to heat the atmosphere above the superburst fuel layer and induce
hydrogen and helium burning, which would power the precursor bursts
observed at the onset of superbursts \citep{Weinberg2007,Keek2011,Keek2012precursors}.
The tail of these precursors and subsequent stable burning of freshly
accreted hydrogen and helium may lower the steepness of the light
curve \citep{Keek2012}. We have reduced this effect by starting our
fits well after the precursor.

\citet{Keek2011} and \citet{Keek2012} calculated fully self-consistent
1d multi-zone models of superbursts. The light curves show both rising
and declining phases at times hundreds to thousands of seconds, depending
on the accretion rates and base fluxes chosen for the models. For
example, the superburst light curve shown in Figure~3 of \citet{Keek2012}
shows a rise between $\approx30$ and $300$ seconds, and then a declining
light curve from $300$ to $10^{4}$ seconds. Our new cooling models
are able to reproduce the light curves of these models. In a future
publication, we will study in detail how the measured temperature
profile is reproduced in these 1d models.

Given that carbon burning to nickel produces $1.0\,\mathrm{MeV\,u^{-1}}=0.96\times10^{18}\,\mathrm{erg\,g^{-1}}$
\citep[ignoring a possible contribution from photodisintegration]{Schatz2003ApJ},
the measured $E_{18}$ translates to a carbon mass fraction of $26\%$.
This is consistent with the value previously measured with local burn
models \citep{Cumming2006}. Furthermore, we find an ignition column
that is less than half as large as previously measured. This is due
to the different shape of our new model light curves. Therefore, the
inferred column depths for other superburst observations may have
been too large as well \citep{Cumming2006,Keek2008,Kuulkers2010,Altamirano2012}.
For fits to the light curve of the candidate superburst from EXO~1745--248,
\citet{Altamirano2012} used an early version of our cooling models
with an $\alpha$ value close to the one that we obtain, in anticipation
of this paper. Because the onset of that superburst was not observed,
it did not itself provide a constraint on $\alpha$. The 2001 superburst
from 4U~1636--536 is the highest quality observation of the rise
and peak, and improved measurements of superburst parameters including
$\alpha$, therefore, require new superburst observations with future
large-area X-ray telescopes such as NASA's \emph{Neutron Star Interior
Composition Explorer} (\emph{NICER}, scheduled for launch in 2016;
\citealt{Gendreau2012NICER}).

\subsection{Accretion history}

Three other superbursts have been observed from 4U~1636--536. The
one immediately preceding the superburst discussed in this paper was
$1.75\,\mathrm{yr}$ earlier (\citealt{Kuulkers2009ATel}; no superburst
has been observed from this source afterwards). From the Multi INstrument
Burst ARchive \citep[MINBAR;][]{Keek2010}, which is the largest catalogue
of X-ray observations of bursting sources, we select all 132 \emph{RXTE}
PCA \citep{Galloway2008catalog} and \emph{BeppoSAX} WFC \citep{Cornelisse2003}
observations in that time interval. The observations are spread out
in time, and constitute a total exposure time of $1.44\,\mathrm{Ms}$.
The unabsorbed persistent $3-25\,\mathrm{keV}$ flux varies between
observations by a factor $2.8$. We determine its time-averaged value
as the mean of all flux measurements weighted by the duration of each
observation: $F_{\mathrm{pers}}=(4.17\pm0.07)\times10^{-9}\,\mathrm{erg\,s^{-1}\,cm^{-2}}$.
Employing a bolometric correction of $1.1$ \citep{Fiocchi2006} and
a distance of $6.0\,\mathrm{kpc}$, the average persistent luminosity
is $L_{\mathrm{pers}}=(1.98\pm0.03)\times10^{37}\,\mathrm{erg\,s^{-1}}$
(distance uncertainty not included). For $1.75$ years of accretion
on to a neutron star with parameters as described in Section~\ref{sub:Details-of-the},
this corresponds to an accretion column of $y_{\mathrm{acc}}=(3.53\pm0.05)\times10^{11}\,\mathrm{g\,cm^{-2}}$,
assuming the efficiency of releasing gravitational potential energy
in X-rays is $100\%$. Because of our assumptions, the actual uncertainty
is probably at least several tens of percent. Given the best-fitting
value for $y_{\mathrm{ign}}$ from cooling models (Section~\ref{sub:Comparison-to-the}),
$y_{\mathrm{acc}}$ can accommodate $2$ superbursts: $y_{\mathrm{acc}}/y_{\mathrm{ign}}=1.8\pm0.2$,
where we use the MCMC-derived error on $y_{\mathrm{ign}}$ without
the uncertainty in the distance, which drops out. We regard it likely
that an extra superburst occurred roughly in the middle of the preceding
$1.75$~years. Around this time no superburst has been detected with,
e.g., the All-Sky Monitor on \emph{RXTE} \citep[e.g.,][]{Kuulkers2009ATel}
or other instruments, but this could easily be attributed to the sparse
coverage of the source by the X-ray observatories that were operational
at the time.

\subsection{Robustness of the observed rise and peak}

In the rise and around the peak of the superburst light curve, the
spectrum clearly separates in three components, one of which we interpret
as the direct thermal emission from the neutron star. The best-fitting
parameters of the considered spectral models for this component can
be somewhat different: e.g., the peak value of $L/L_{\mathrm{Edd}}$
for the metal-poor atmosphere model is $12\%$ lower than for the
solar-composition atmosphere. Nevertheless, the neutron star flux
is consistent between the spectral models of both a blackbody and
neutron star atmospheres. All models result in a similar goodness
of fit.

After the peak, at $t\simeq1400\,\mathrm{s}$  the light curve exhibits
some variability on time-scales of $\sim10^{2}\,\mathrm{s}$. Compared
to the cooling model fit to the data at earlier times, the flux variations
are both above and below the model, and may have a similar origin
as the so-called ``achromatic'' variability, which is typically
observed in some Eddington-limited bursts including the 1999 superburst
from 4U~1820--30 \citep{Zand2011}. In our spectral fits, the variability
is strongest in the neutron star component. Since we constrained the
persistent component to be the same throughout the first orbit (Section~\ref{sub:Spectral-Analysis}),
however, it is possible that the variability is actually part of the
persistent emission. Instead of the neutron star envelope, it may
be more likely that the physical origin of the variability is found
in, e.g., instabilities in the accretion disc.

A systematic uncertainty in the conversion of flux to luminosity is
introduced by anisotropy of the emission. Assuming an inclination
angle of $60^{\circ}$ \citep[e.g.,][]{Pandel2008} and a thin flat
disc, the neutron star's intrinsic luminosity is $\nicefrac{1}{3}$
larger than the value we obtained under the assumption of isotropic
emission \citep{fujimoto88apj}. Its main effect is an equivalent
increase in $E_{18}$.

The most robust part of the superburst light curve is, therefore,
the rise and peak, until the onset of variability. We, therefore,
fit cooling models exclusively to this part.

\subsection{Interpretation of the spectra in the tail}

\label{sub:Interpretation-of-the}

In the tail, the observed signal is weaker, and it is challenging
to separate and interpret the spectral components. Especially the
spectral fits with atmosphere models are problematic, as at certain
times the fits cannot distinguish between very low values of $L/L_{\mathrm{Edd}}$
and more reasonable values (Figure~\ref{fig:compare-flux}). Even
when disregarding these unphysical solutions, for all spectral models
the neutron star envelope's temperature appears to drop faster than
predicted by cooling models (Figure~\ref{fig:alpha}). \citet{Schatz2014Nature}
pointed out that an URCA neutrino cooling process could operate in
the outer crust, depending on the mixture of nuclei present, and suggested
that the cooling could affect the shape of superburst tails. However,
the drop in temperature that we observe here is much faster than expected
even with URCA cooling. The disagreement between the model and data
occurs immediately after the peak, implying that the URCA cooling
source would have to be at a very shallow depth close to the carbon
ignition depth, which is unlikely. Furthermore, fully self-consistent
1d models of superbursts have similar problems in simultaneously fitting
the rise and tail \citep[albeit this comparison was against a lightcurve that did not account for increased persistent emission nor reflection]{Keek2011},
indicating that the mismatch is not due to, e.g., late-time residual
burning. The strong temperature drop is found for the different constraints
on the normalization of the neutron star spectral component (BB and
$\mathrm{BB_{\mathrm{frac}}}$) as well as for the combined neutron
star and reflection components \citep{Keek2014sb1}. If the neutron
star's emitting area is kept constant, this directly translates into
a fast decline of $F_{\mathrm{NS}}$ (BB), whereas with a constant
reflection fraction $F_{\mathrm{NS}}$ more closely matches the best-fitting
cooling model ($\mathrm{BB_{\mathrm{frac}}}$; Figure~\ref{fig:compare-model}).

The slower decay of $F_{\mathrm{NS}}$ for $\mathrm{BB_{\mathrm{frac}}}$
requires a radius increase of a factor $2.27\pm0.04$ in the tail,
even though radius expansion is typically expected to occur only at
larger luminosities near the Eddington limit. Alternatively, deviations
of the neutron star spectrum from a blackbody have manifested itself
as increased normalizations in the tail of short X-ray bursts \citep[e.g.,][]{Suleimanov2010}.
By fitting with atmosphere models (S1 and S001), we confirm that this
plays no role in our case, because we fix the normalization in the
tail. Furthermore, the neutron star spectral component is consistent
with a single-temperature blackbody, indicating that the fast temperature
decrease is intrinsic to the neutron star. In this case the radius
increase must be intrinsic to the star as well, and not an anisotropy
effect \citep{lapidus85mnras,fujimoto88apj}. 

Finally, as the flux in the tail is dominated by the persistent component
(Figure~\ref{fig:compare-flux}; see also \citealt{Keek2014sb1,Keek2014sb2}),
mischaracterisation of this component could influence the results
of the analysis in this part of the superburst. For example, a boundary
or spreading layer is thought to be present between the accretion
disc and the neutron star atmosphere \citep{Inogamov1999}. The fraction
of the neutron star covered by this layer may evolve during the superburst.
It is, however, challenging to decompose even high-quality spectra
of persistent emission \citep{Revnivtsev2013}. Better constraints
on the persistent spectrum may provided by a future superburst observation
with energy coverage extending below $1\,\mathrm{keV}$, for example
with \emph{XMM Newton} \citep{jansen01aa} or \emph{NICER} \citep{Gendreau2012NICER}.

\section{Conclusions}

The 2001 \emph{RXTE}/PCA superburst from 4U~1636--536 is the only
high-quality observation of the intrinsic rise and peak of a superburst.
We separate the direct thermal emission from the neutron star from
evolving reflection and persistent contributions to the spectrum,
and evaluate the robustness of the derived light curve with respect
to the choice of spectral model, including neutron star atmosphere
models. In the tail of the burst, where the count rate is lowest,
the interpretation of the spectrum is ambiguous. The rise and peak,
however, provide a robust net superburst light curve. We show how
the steepness of the rise is determined by the slope of the temperature
profile in the neutron star envelope. The light curve is fit with
new cooling models that take this feature into account. The temperature
slope with pressure is constrained to be $T\propto P^{\alpha}$ with
$\alpha\approx0.25$. This is inconsistent with complete burning of
the carbon to iron group elements, and implies significant incomplete
burning. Theoretical calculations of the propagation of carbon flames
in neutron star envelopes are needed to explain the observed slope.
Finally, the observational upper limit to the recurrence time of this
superburst is $1.75\,\mathrm{yr}$, but we find that in that time
enough fuel was accumulated for two superbursts. This implies that
the superburst directly prior to the one in 2001 had escaped detection.

\section*{Acknowledgements}

We are grateful for discussions with E.\,F.~Brown, A.~Heger, and
D.\,K.~Galloway. LK and DRB acknowledge support from NASA ADAP grant
NNX13AI47G and NSF award AST 1008067. AC was supported by the National
Sciences and Engineering Research Council (NSERC) of Canada. AC is
an associate of the CIFAR Cosmology and Gravity program. VS was supported
by German Research Foundation (DFG) grant WE 1312/48-1. This paper
uses preliminary analysis results from the Multi-INstrument Burst
ARchive (MINBAR), which is supported under the Australian Academy
of Science's Scientific Visits to Europe program, and the Australian
Research Council's Discovery Projects and Future Fellowship funding
schemes. We made use of the emcee code by \citeauthor{ForemanMackey2013emcee},
available at \url{http://dan.iel.fm/emcee}.

\bibliographystyle{mnras}
\bibliography{sbatmo}

\begin{thebibliography}{}
\makeatletter
\relax
\def\mn@urlcharsother{\let\do\@makeother \do\$\do\&\do\#\do\^\do\_\do\%\do\~}
\def\mn@doi{\begingroup\mn@urlcharsother \@ifnextchar [ {\mn@doi@}
  {\mn@doi@[]}}
\def\mn@doi@[#1]#2{\def\@tempa{#1}\ifx\@tempa\@empty \href
  {http://dx.doi.org/#2} {doi:#2}\else \href {http://dx.doi.org/#2} {#1}\fi
  \endgroup}
\def\mn@eprint#1#2{\mn@eprint@#1:#2::\@nil}
\def\mn@eprint@arXiv#1{\href {http://arxiv.org/abs/#1} {{\tt arXiv:#1}}}
\def\mn@eprint@dblp#1{\href {http://dblp.uni-trier.de/rec/bibtex/#1.xml}
  {dblp:#1}}
\def\mn@eprint@#1:#2:#3:#4\@nil{\def\@tempa {#1}\def\@tempb {#2}\def\@tempc
  {#3}\ifx \@tempc \@empty \let \@tempc \@tempb \let \@tempb \@tempa \fi \ifx
  \@tempb \@empty \def\@tempb {arXiv}\fi \@ifundefined
  {mn@eprint@\@tempb}{\@tempb:\@tempc}{\expandafter \expandafter \csname
  mn@eprint@\@tempb\endcsname \expandafter{\@tempc}}}

\bibitem[\protect\citeauthoryear{{Altamirano} et~al.,}{{Altamirano}
  et~al.}{2012}]{Altamirano2012}
{Altamirano} D.,  et~al., 2012, \mn@doi [\mnras]
  {10.1111/j.1365-2966.2012.21769.x},
  \href{http://cdsads.u-strasbg.fr/abs/2012MNRAS.426..927A}{426, 927}

\bibitem[\protect\citeauthoryear{{Arnaud}}{{Arnaud}}{1996}]{Arnaud1996}
{Arnaud} K.~A.,  1996, in {Jacoby} G.~H.,  {Barnes} J.,  eds,  Astronomical
  Society of the Pacific Conference Series Vol. 101, Astronomical Data Analysis
  Software and Systems V. p.~17

\bibitem[\protect\citeauthoryear{{Ballantyne}}{{Ballantyne}}{2004}]{Ballantyne2004models}
{Ballantyne} D.~R.,  2004, \mn@doi [\mnras] {10.1111/j.1365-2966.2004.07767.x},
  \href{http://adsabs.harvard.edu/abs/2004MNRAS.351...57B}{351, 57}

\bibitem[\protect\citeauthoryear{{Ballantyne} \& {Everett}}{{Ballantyne} \&
  {Everett}}{2005}]{Ballantyne2005}
{Ballantyne} D.~R.,  {Everett} J.~E.,  2005, \mn@doi [\apj] {10.1086/429860},
  \href{http://adsabs.harvard.edu/abs/2005ApJ...626..364B}{626, 364}

\bibitem[\protect\citeauthoryear{{Ballantyne} \& {Strohmayer}}{{Ballantyne} \&
  {Strohmayer}}{2004}]{Ballantyne2004}
{Ballantyne} D.~R.,  {Strohmayer} T.~E.,  2004, \mn@doi [\apjl]
  {10.1086/382703},
  \href{http://adsabs.harvard.edu/abs/2004ApJ...602L.105B}{602, L105}

\bibitem[\protect\citeauthoryear{{Ballantyne}, {Fabian}  \&
  {Ross}}{{Ballantyne} et~al.}{2002}]{Ballantyne2002code}
{Ballantyne} D.~R.,  {Fabian} A.~C.,   {Ross} R.~R.,  2002, \mn@doi [\mnras]
  {10.1046/j.1365-8711.2002.05241.x},
  \href{http://adsabs.harvard.edu/abs/2002MNRAS.329L..67B}{329, L67}

\bibitem[\protect\citeauthoryear{{Bradt}, {Rothschild}  \& {Swank}}{{Bradt}
  et~al.}{1993}]{Bradt1993}
{Bradt} H.~V.,  {Rothschild} R.~E.,   {Swank} J.~H.,  1993, \aaps,
  \href{http://adsabs.harvard.edu/abs/1993A%26AS...97..355B}{97, 355}

\bibitem[\protect\citeauthoryear{{Chabrier} \& {Potekhin}}{{Chabrier} \&
  {Potekhin}}{1998}]{Chabrier1998}
{Chabrier} G.,  {Potekhin} A.~Y.,  1998, \mn@doi [\pre]
  {10.1103/PhysRevE.58.4941},
  \href{http://adsabs.harvard.edu/abs/1998PhRvE..58.4941C}{58, 4941}

\bibitem[\protect\citeauthoryear{{Cornelisse}, {Heise}, {Kuulkers}, {Verbunt}
  \& {in't Zand}}{{Cornelisse} et~al.}{2000}]{Cornelisse2000}
{Cornelisse} R.,  {Heise} J.,  {Kuulkers} E.,  {Verbunt} F.,   {in't Zand}
  J.~J.~M.,  2000, \aap,
  \href{http://adsabs.harvard.edu/abs/2000A%26A...357L..21C}{357, L21}

\bibitem[\protect\citeauthoryear{{Cornelisse} et~al.,}{{Cornelisse}
  et~al.}{2003}]{Cornelisse2003}
{Cornelisse} R.,  et~al., 2003, \mn@doi [\aap] {10.1051/0004-6361:20030629},
  \href{http://adsabs.harvard.edu/abs/2003A%26A...405.1033C}{405, 1033}

\bibitem[\protect\citeauthoryear{{Cumming} \& {Bildsten}}{{Cumming} \&
  {Bildsten}}{2001}]{Cumming2001}
{Cumming} A.,  {Bildsten} L.,  2001, \apjl,
  \href{http://cdsads.u-strasbg.fr/abs/2001ApJ...559L.127C}{559, L127}

\bibitem[\protect\citeauthoryear{{Cumming} \& {Macbeth}}{{Cumming} \&
  {Macbeth}}{2004}]{2004CummingMacBeth}
{Cumming} A.,  {Macbeth} J.,  2004, \mn@doi [\apjl] {10.1086/382873},
  \href{http://adsabs.harvard.edu/cgi-bin/nph-bib_query?bibcode=2004ApJ...603L..37C&db_key=AST}{603,
  L37}

\bibitem[\protect\citeauthoryear{{Cumming}, {Macbeth}, {in~'t~Zand}  \&
  {Page}}{{Cumming} et~al.}{2006}]{Cumming2006}
{Cumming} A.,  {Macbeth} J.,  {in~'t~Zand} J.~J.~M.,   {Page} D.,  2006,
  \mn@doi [\apj] {10.1086/504698},
  \href{http://adsabs.harvard.edu/abs/2006ApJ...646..429C}{646, 429}

\bibitem[\protect\citeauthoryear{{Fabian}, {Rees}, {Stella}  \&
  {White}}{{Fabian} et~al.}{1989}]{Fabian1989}
{Fabian} A.~C.,  {Rees} M.~J.,  {Stella} L.,   {White} N.~E.,  1989, \mnras,
  \href{http://adsabs.harvard.edu/abs/1989MNRAS.238..729F}{238, 729}

\bibitem[\protect\citeauthoryear{{Fiocchi}, {Bazzano}, {Ubertini}  \&
  {Jean}}{{Fiocchi} et~al.}{2006}]{Fiocchi2006}
{Fiocchi} M.,  {Bazzano} A.,  {Ubertini} P.,   {Jean} P.,  2006, \mn@doi [\apj]
  {10.1086/507321},
  \href{http://adsabs.harvard.edu/abs/2006ApJ...651..416F}{651, 416}

\bibitem[\protect\citeauthoryear{{Foreman-Mackey}, {Hogg}, {Lang}  \&
  {Goodman}}{{Foreman-Mackey} et~al.}{2013}]{ForemanMackey2013emcee}
{Foreman-Mackey} D.,  {Hogg} D.~W.,  {Lang} D.,   {Goodman} J.,  2013, \mn@doi
  [\pasp] {10.1086/670067},
  \href{http://adsabs.harvard.edu/abs/2013PASP..125..306F}{125, 306}

\bibitem[\protect\citeauthoryear{{Fujimoto}}{{Fujimoto}}{1988}]{fujimoto88apj}
{Fujimoto} M.~Y.,  1988, \mn@doi [\apj] {10.1086/165955},
  \href{http://adsabs.harvard.edu/abs/1988ApJ...324..995F}{324, 995}

\bibitem[\protect\citeauthoryear{{Galloway}, {Muno}, {Hartman}, {Psaltis}  \&
  {Chakrabarty}}{{Galloway} et~al.}{2008}]{Galloway2008catalog}
{Galloway} D.~K.,  {Muno} M.~P.,  {Hartman} J.~M.,  {Psaltis} D.,
  {Chakrabarty} D.,  2008, \mn@doi [\apjs] {10.1086/592044},
  \href{http://adsabs.harvard.edu/abs/2008ApJS..179..360G}{179, 360}

\bibitem[\protect\citeauthoryear{{Gendreau}, {Arzoumanian}  \&
  {Okajima}}{{Gendreau} et~al.}{2012}]{Gendreau2012NICER}
{Gendreau} K.~C.,  {Arzoumanian} Z.,   {Okajima} T.,  2012, in Society of
  Photo-Optical Instrumentation Engineers (SPIE) Conference Series. ,
  \mn@doi{10.1117/12.926396}

\bibitem[\protect\citeauthoryear{{Haensel}, {Kaminker}  \&
  {Yakovlev}}{{Haensel} et~al.}{1996}]{Haensel1996}
{Haensel} P.,  {Kaminker} A.~D.,   {Yakovlev} D.~G.,  1996, \aap,
  \href{http://adsabs.harvard.edu/abs/1996A%26A...314..328H}{314, 328}

\bibitem[\protect\citeauthoryear{{Inogamov} \& {Sunyaev}}{{Inogamov} \&
  {Sunyaev}}{1999}]{Inogamov1999}
{Inogamov} N.~A.,  {Sunyaev} R.~A.,  1999, Astronomy Letters,
  \href{http://adsabs.harvard.edu/abs/1999AstL...25..269I}{25, 269}

\bibitem[\protect\citeauthoryear{{Jahoda}, {Markwardt}, {Radeva}, {Rots},
  {Stark}, {Swank}, {Strohmayer}  \& {Zhang}}{{Jahoda}
  et~al.}{2006}]{Jahoda2006}
{Jahoda} K.,  {Markwardt} C.~B.,  {Radeva} Y.,  {Rots} A.~H.,  {Stark} M.~J.,
  {Swank} J.~H.,  {Strohmayer} T.~E.,   {Zhang} W.,  2006, \mn@doi [\apjs]
  {10.1086/500659},
  \href{http://adsabs.harvard.edu/abs/2006ApJS..163..401J}{163, 401}

\bibitem[\protect\citeauthoryear{{Jansen} et~al.,}{{Jansen}
  et~al.}{2001}]{jansen01aa}
{Jansen} F.,  et~al., 2001, \aap, 365, L1

\bibitem[\protect\citeauthoryear{{Keek}}{{Keek}}{2012}]{Keek2012precursors}
{Keek} L.,  2012, \mn@doi [\apj] {10.1088/0004-637X/756/2/130},
  \href{http://adsabs.harvard.edu/abs/2012ApJ...756..130K}{756, 130}

\bibitem[\protect\citeauthoryear{{Keek} \& {Heger}}{{Keek} \&
  {Heger}}{2011}]{Keek2011}
{Keek} L.,  {Heger} A.,  2011, \mn@doi [\apj] {10.1088/0004-637X/743/2/189},
  \href{http://adsabs.harvard.edu/abs/2011ApJ...743..189K}{743, 189}

\bibitem[\protect\citeauthoryear{{Keek}, {in~'t~Zand}, {Kuulkers}, {Cumming},
  {Brown}  \& {Suzuki}}{{Keek} et~al.}{2008}]{Keek2008}
{Keek} L.,  {in~'t~Zand} J.~J.~M.,  {Kuulkers} E.,  {Cumming} A.,  {Brown}
  E.~F.,   {Suzuki} M.,  2008, \mn@doi [\aap] {10.1051/0004-6361:20078464},
  \href{http://adsabs.harvard.edu/abs/2008A%26A...479..177K}{479, 177}

\bibitem[\protect\citeauthoryear{{Keek}, {Galloway}, {in 't Zand}  \&
  {Heger}}{{Keek} et~al.}{2010}]{Keek2010}
{Keek} L.,  {Galloway} D.~K.,  {in 't Zand} J.~J.~M.,   {Heger} A.,  2010,
  \mn@doi [\apj] {10.1088/0004-637X/718/1/292},
  \href{http://adsabs.harvard.edu/abs/2010ApJ...718..292K}{718, 292}

\bibitem[\protect\citeauthoryear{{Keek}, {Heger}  \& {in't Zand}}{{Keek}
  et~al.}{2012}]{Keek2012}
{Keek} L.,  {Heger} A.,   {in't Zand} J.~J.~M.,  2012, \mn@doi [\apj]
  {10.1088/0004-637X/752/2/150},
  \href{http://adsabs.harvard.edu/abs/2012ApJ...752..150K}{752, 150}

\bibitem[\protect\citeauthoryear{{Keek}, {Ballantyne}, {Kuulkers}  \&
  {Strohmayer}}{{Keek} et~al.}{2014a}]{Keek2014sb1}
{Keek} L.,  {Ballantyne} D.~R.,  {Kuulkers} E.,   {Strohmayer} T.~E.,  2014a,
  \mn@doi [\apj] {10.1088/0004-637X/789/2/121},
  \href{http://adsabs.harvard.edu/abs/2014ApJ...789..121K}{789, 121}

\bibitem[\protect\citeauthoryear{{Keek}, {Ballantyne}, {Kuulkers}  \&
  {Strohmayer}}{{Keek} et~al.}{2014b}]{Keek2014sb2}
{Keek} L.,  {Ballantyne} D.~R.,  {Kuulkers} E.,   {Strohmayer} T.~E.,  2014b,
  \mn@doi [\apjl] {10.1088/2041-8205/797/2/L23},
  \href{http://adsabs.harvard.edu/abs/2014ApJ...797L..23K}{797, L23}

\bibitem[\protect\citeauthoryear{{Kuulkers}}{{Kuulkers}}{2009}]{Kuulkers2009ATel}
{Kuulkers} E.,  2009, The Astronomer's Telegram,
  \href{http://adsabs.harvard.edu/abs/2009ATel.2140....1K}{2140, 1}

\bibitem[\protect\citeauthoryear{{Kuulkers}, {in't Zand}, {Homan}, {van
  Straaten}, {Altamirano}  \& {van der Klis}}{{Kuulkers}
  et~al.}{2004}]{2004Kuulkers}
{Kuulkers} E.,  {in't Zand} J.,  {Homan} J.,  {van Straaten} S.,  {Altamirano}
  D.,   {van der Klis} M.,  2004, in {Kaaret} P.,  {Lamb} F.~K.,   {Swank}
  J.~H.,  eds,  American Institute of Physics Conference Series Vol. 714, X-ray
  Timing 2003: Rossi and Beyond. pp 257--260 (\mn@eprint {}
  {astro-ph/0402076}), \mn@doi{10.1063/1.1781037}

\bibitem[\protect\citeauthoryear{{Kuulkers} et~al.,}{{Kuulkers}
  et~al.}{2010}]{Kuulkers2010}
{Kuulkers} E.,  et~al., 2010, \mn@doi [\aap] {10.1051/0004-6361/200913210},
  \href{http://adsabs.harvard.edu/abs/2010A%26A...514A..65K}{514, A65+}

\bibitem[\protect\citeauthoryear{{Lapidus} \& {Sunyaev}}{{Lapidus} \&
  {Sunyaev}}{1985}]{lapidus85mnras}
{Lapidus} I.~I.,  {Sunyaev} R.~A.,  1985, \mnras,
  \href{http://adsabs.harvard.edu/abs/1985MNRAS.217..291L}{217, 291}

\bibitem[\protect\citeauthoryear{{Lewin}, {van Paradijs}  \& {Taam}}{{Lewin}
  et~al.}{1993}]{Lewin1993}
{Lewin} W.~H.~G.,  {van Paradijs} J.,   {Taam} R.~E.,  1993, \mn@doi [Space
  Science Reviews] {10.1007/BF00196124},
  \href{http://cdsads.u-strasbg.fr/abs/1993SSRv...62..223L}{62, 223}

\bibitem[\protect\citeauthoryear{{Negoro} et~al.,}{{Negoro}
  et~al.}{2012}]{Negoro2012}
{Negoro} H.,  et~al., 2012, The Astronomer's Telegram,
  \href{http://adsabs.harvard.edu/abs/2012ATel.4622....1N}{4622, 1}

\bibitem[\protect\citeauthoryear{{Paczynski}}{{Paczynski}}{1983}]{Paczynski1983}
{Paczynski} B.,  1983, \mn@doi [\apj] {10.1086/160596},
  \href{http://adsabs.harvard.edu/abs/1983ApJ...264..282P}{264, 282}

\bibitem[\protect\citeauthoryear{{Pandel}, {Kaaret}  \& {Corbel}}{{Pandel}
  et~al.}{2008}]{Pandel2008}
{Pandel} D.,  {Kaaret} P.,   {Corbel} S.,  2008, \mn@doi [\apj]
  {10.1086/592429},
  \href{http://adsabs.harvard.edu/abs/2008ApJ...688.1288P}{688, 1288}

\bibitem[\protect\citeauthoryear{{Potekhin} \& {Chabrier}}{{Potekhin} \&
  {Chabrier}}{2000}]{Potekhin2000}
{Potekhin} A.~Y.,  {Chabrier} G.,  2000, \mn@doi [\pre]
  {10.1103/PhysRevE.62.8554},
  \href{http://adsabs.harvard.edu/abs/2000PhRvE..62.8554P}{62, 8554}

\bibitem[\protect\citeauthoryear{{Potekhin} \& {Yakovlev}}{{Potekhin} \&
  {Yakovlev}}{2001}]{Potekhin2001}
{Potekhin} A.~Y.,  {Yakovlev} D.~G.,  2001, \mn@doi [\aap]
  {10.1051/0004-6361:20010698},
  \href{http://adsabs.harvard.edu/abs/2001A%26A...374..213P}{374, 213}

\bibitem[\protect\citeauthoryear{{Potekhin}, {Baiko}, {Haensel}  \&
  {Yakovlev}}{{Potekhin} et~al.}{1999}]{Potekhin1999}
{Potekhin} A.~Y.,  {Baiko} D.~A.,  {Haensel} P.,   {Yakovlev} D.~G.,  1999,
  \aap, \href{http://adsabs.harvard.edu/abs/1999A%26A...346..345P}{346, 345}

\bibitem[\protect\citeauthoryear{{Revnivtsev}, {Suleimanov}  \&
  {Poutanen}}{{Revnivtsev} et~al.}{2013}]{Revnivtsev2013}
{Revnivtsev} M.~G.,  {Suleimanov} V.~F.,   {Poutanen} J.,  2013, \mn@doi
  [\mnras] {10.1093/mnras/stt1179},
  \href{http://adsabs.harvard.edu/abs/2013MNRAS.434.2355R}{434, 2355}

\bibitem[\protect\citeauthoryear{{Schatz}, {Bildsten}, {Cumming}  \&
  {Wiescher}}{{Schatz} et~al.}{1999}]{1999Schatz}
{Schatz} H.,  {Bildsten} L.,  {Cumming} A.,   {Wiescher} M.,  1999, \mn@doi
  [\apj] {10.1086/307837},
  \href{http://adsabs.harvard.edu/cgi-bin/nph-bib_query?bibcode=1999ApJ...524.1014S&db_key=AST}{524,
  1014}

\bibitem[\protect\citeauthoryear{{Schatz}, {Bildsten}  \& {Cumming}}{{Schatz}
  et~al.}{2003}]{Schatz2003ApJ}
{Schatz} H.,  {Bildsten} L.,   {Cumming} A.,  2003, \mn@doi [\apjl]
  {10.1086/368107},
  \href{http://adsabs.harvard.edu/abs/2003ApJ...583L..87S}{583, L87}

\bibitem[\protect\citeauthoryear{{Schatz} et~al.,}{{Schatz}
  et~al.}{2014}]{Schatz2014Nature}
{Schatz} H.,  et~al., 2014, \mn@doi [\nat] {10.1038/nature12757},
  \href{http://adsabs.harvard.edu/abs/2014Natur.505...62S}{505, 62}

\bibitem[\protect\citeauthoryear{{Schinder}, {Schramm}, {Wiita}, {Margolis}  \&
  {Tubbs}}{{Schinder} et~al.}{1987}]{Schinder1987}
{Schinder} P.~J.,  {Schramm} D.~N.,  {Wiita} P.~J.,  {Margolis} S.~H.,
  {Tubbs} D.~L.,  1987, \mn@doi [\apj] {10.1086/164993},
  \href{http://adsabs.harvard.edu/abs/1987ApJ...313..531S}{313, 531}

\bibitem[\protect\citeauthoryear{{Serino} et~al.,}{{Serino}
  et~al.}{2014}]{Serino2014ATel}
{Serino} M.,  et~al., 2014, The Astronomer's Telegram,
  \href{http://adsabs.harvard.edu/abs/2014ATel.6668....1S}{6668, 1}

\bibitem[\protect\citeauthoryear{{Stevens}, {Brown}, {Cumming}, {Cyburt}  \&
  {Schatz}}{{Stevens} et~al.}{2014}]{Stevens2014}
{Stevens} J.,  {Brown} E.~F.,  {Cumming} A.,  {Cyburt} R.,   {Schatz} H.,
  2014, \mn@doi [\apj] {10.1088/0004-637X/791/2/106},
  \href{http://adsabs.harvard.edu/abs/2014ApJ...791..106S}{791, 106}

\bibitem[\protect\citeauthoryear{{Strohmayer} \& {Brown}}{{Strohmayer} \&
  {Brown}}{2002}]{Strohmayer2002}
{Strohmayer} T.~E.,  {Brown} E.~F.,  2002, \mn@doi [\apj] {10.1086/338337},
  \href{http://adsabs.harvard.edu/abs/2002ApJ...566.1045S}{566, 1045}

\bibitem[\protect\citeauthoryear{{Strohmayer} \& {Markwardt}}{{Strohmayer} \&
  {Markwardt}}{2002}]{Strohmayer2002a}
{Strohmayer} T.~E.,  {Markwardt} C.~B.,  2002, \mn@doi [\apj] {10.1086/342152},
  \href{http://adsabs.harvard.edu/abs/2002ApJ...577..337S}{577, 337}

\bibitem[\protect\citeauthoryear{{Suleimanov}, {Poutanen}  \&
  {Werner}}{{Suleimanov} et~al.}{2011}]{Suleimanov2010}
{Suleimanov} V.,  {Poutanen} J.,   {Werner} K.,  2011, \mn@doi [\aap]
  {10.1051/0004-6361/201015845},
  \href{http://adsabs.harvard.edu/abs/2011A%26A...527A.139S}{527, A139+}

\bibitem[\protect\citeauthoryear{{Suleimanov}, {Poutanen}  \&
  {Werner}}{{Suleimanov} et~al.}{2012}]{Suleimanov2012}
{Suleimanov} V.,  {Poutanen} J.,   {Werner} K.,  2012, \mn@doi [\aap]
  {10.1051/0004-6361/201219480},
  \href{http://adsabs.harvard.edu/abs/2012A%26A...545A.120S}{545, A120}

\bibitem[\protect\citeauthoryear{{Weinberg} \& {Bildsten}}{{Weinberg} \&
  {Bildsten}}{2007}]{Weinberg2007}
{Weinberg} N.~N.,  {Bildsten} L.,  2007, \mn@doi [\apj] {10.1086/522111},
  \href{http://adsabs.harvard.edu/abs/2007ApJ...670.1291W}{670, 1291}

\bibitem[\protect\citeauthoryear{{Woosley}, {Kerstein}  \& {Aspden}}{{Woosley}
  et~al.}{2011}]{Woosley2011}
{Woosley} S.~E.,  {Kerstein} A.~R.,   {Aspden} A.~J.,  2011, \mn@doi [\apj]
  {10.1088/0004-637X/734/1/37},
  \href{http://adsabs.harvard.edu/abs/2011ApJ...734...37W}{734, 37}

\bibitem[\protect\citeauthoryear{{in~'t Zand} \& {Weinberg}}{{in~'t Zand} \&
  {Weinberg}}{2010}]{Zand2010}
{in~'t Zand} J.~J.~M.,  {Weinberg} N.~N.,  2010, \mn@doi [\aap]
  {10.1051/0004-6361/200913952},
  \href{http://esoads.eso.org/abs/2010A%26A...520A..81I}{520, A81}

\bibitem[\protect\citeauthoryear{{in~'t~Zand}, {Kuulkers}, {Verbunt}, {Heise}
  \& {Cornelisse}}{{in~'t~Zand} et~al.}{2003}]{Zand2003}
{in~'t~Zand} J.~J.~M.,  {Kuulkers} E.,  {Verbunt} F.,  {Heise} J.,
  {Cornelisse} R.,  2003, \mn@doi [\aap] {10.1051/0004-6361:20031586},
  \href{http://adsabs.harvard.edu/abs/2003A%26A...411L.487I}{411, L487}

\bibitem[\protect\citeauthoryear{{in~'t~Zand}, {Cornelisse}  \&
  {Cumming}}{{in~'t~Zand} et~al.}{2004}]{Zand2004}
{in~'t~Zand} J.~J.~M.,  {Cornelisse} R.,   {Cumming} A.,  2004, \mn@doi [\aap]
  {10.1051/0004-6361:20040522},
  \href{http://adsabs.harvard.edu/abs/2004A%26A...426..257I}{426, 257}

\bibitem[\protect\citeauthoryear{{in~'t~Zand}, {Galloway}  \&
  {Ballantyne}}{{in~'t~Zand} et~al.}{2011}]{Zand2011}
{in~'t~Zand} J.~J.~M.,  {Galloway} D.~K.,   {Ballantyne} D.~R.,  2011, \mn@doi
  [\aap] {10.1051/0004-6361/201015556},
  \href{http://adsabs.harvard.edu/abs/2011A%26A...525A.111I}{525, A111}

\makeatother
\end{thebibliography}

\bsp	
\label{lastpage}
\end{document}